# The Full Nonlinear Vortex Tube-Vorton Method: the post-stall condition

Jesús Carlos Pimentel-García[1], and Luis Antonio Jiménez-Ponce[2]

[1] Independent researcher, Mexico

[2] Aeronautical engineering department, Technological Institute of Higher Studies of Ecatepec (TESE), 55210 Mexico

Correspondence: [1] pimentel_garcia@yahoo.com.mx, [2] lajimenezp@tese.edu.mx

## Abstract

The original concepts behind *The Full Nonlinear Vortex Tube-Vorton Method* (FTVM) have been applied to the study of massively separated fluid flow past a thin body. In the pre-stall condition, the Kutta-Zhukovski (KJ) force calculation was successfully implemented. However, for the post-stall condition, this approach is not capable of capturing the form drag component. Thus, an alternative force calculation method is implemented. The results obtained demonstrate that the fluid motion involving complex wake dynamics can be accurately approximated by a low-order, low-discretization numerical method based purely on vorticity and velocity concepts. This method avoids the need to discuss and speculate about the role of pressure on fluid motion, even in the turbulent regime. Consequently, it simplifies the understanding of complex low-speed aerodynamics by returning to the first principles.

**Keywords:** Vortex methods, separated fluid, post-stall regime, stalled flat plate, low-aspect ratio aerodynamics, unsteady aerodynamic force, turbulent vortex wakes

## 1 Introduction

Simulating massively separated fluid flow past a three-dimensional object is a significant challenge in computational fluid dynamics (CFD). This difficulty arises because such flows inherently exhibit turbulent behavior, characterized by *chaotic changes in pressure and flow velocity*, and involve phenomena as mixing and decay of vortex wakes downstream.

In the context of mesh-based solutions, such as Reynolds-averaged Navier-Stokes (RANS), large eddy (LES), and direct numerical simulations (DNS), there are two fundamental approaches for approximating turbulence: modeling and *solving* the turbulent scales. Turbulence models are not more than semi-empirical approximations, which depend on a large number of coefficients to adjust functions, depending on the operating and boundary conditions of a particular simulation. For instance, one of the simplest turbulence models, the original one-equation Spalart-Allmaras [1], requires at least eight constants to calibrate the transport equation for the kinematic eddy viscosity. All of these constants are arbitrarily assigned based on recommendations and modifying them demands a more profound comprehension of the model's functionality and its physical implications. This can potentially lead to simulations being performed as a *black box* by non-experienced users, who may then assume that the obtained numerical results represent real physics of the problem.

As anticipated, the implementation of more sophisticated turbulence models (e.g., shear stress transport; SST) involves a more profound comprehension. The selection of appropriate constants is thus entrusted to experts in the fields of turbulence modeling and experimental testing. For the majority of users, this process often devolves into a trial-and-error endeavor, resulting in the reliance on guesswork until numerical results align *more or less* with the expectations. It must be noted that this approach should not be supported by the standards of formal science; however, it is frequently employed for practical or *engineering purposes*.

In order to address this issue, a proposed solution involves the direct resolution of all turbulent scales by employing a mesh-based alternative: the DNS. Despite the advancements in computational efficiency, enabled by modern hardware, algorithms, and techniques (e.g., GPU parallelization), this approach continues to demand a substantial computational effort and energy consumption. Consequently, obtaining such solutions remains impractical, even within the context of engineering applications. However, DNS indicates that the most straightforward and effective approach for resolving fluid flow phenomena is ideally a parameter-free strategy for approximating fluid dynamics, including turbulence. Nonetheless, it is unlikely that this approach can be implemented efficiently in the near future. Meanwhile, LES, an intermediate approach between RANS and DNS, is gradually replacing the average-based modeling.

In contrast, potential flow-based solutions (e.g., panel, vortex lattice, and lifting line methods) are characterized by their low computational cost, as they do not require three-dimensional meshes. These methods have a long history of application in approximating, above all, simple aerodynamic simulations under ideal conditions by solving the Laplace equation ($\nabla^2\varphi = 0$), the most simplified version of the equations of fluid motion.

The attached flow assumption of potential-based solutions exhibits compatibility with high-Reynolds number (Re) cases, given that, from a physical perspective, the viscous boundary layer (BL) is confined to a thin region in proximity to the surface. This phenomenon can be modeled through the implementation of a vorticity distribution on the surface, employing techniques such as the vortex panel (thick body) and the vortex lattice (thin body) methods, in two or three dimensions. From this perspective, ideal flows in the context of the potential flow theory (PFT) have historically been defined as inviscid ($\nu = 0$), incompressible ($\nabla \cdot v = 0$), and irrotational ($\omega = 0$). However, since vorticity has been demonstrated to be present at the surface, the irrotationality condition is not strictly satisfied everywhere. This means that vorticity exists, even in a supposedly *inviscid* frame.

In this sense, it must be acknowledged that the notion of attached or "*bound vorticity"* remains vague. Essentially, vorticity implies flow separation. In other words, a vortex cannot coexist with a solid surface because vorticity is annihilated nearby, including the inviscid blocking phenomenon. In the viscous case, this diffusion effect is mainly attributed to the shear stress mechanism in proximity to a wall. In the context of current potential-based solutions, it is imperative to acknowledge the existence of flow detachment, even in the limit of an infinitely thin BL, as depicted by a vortex sheet [2], subsequently advected downstream from a prescribed separation point (2D) or line (3D), formally referred to as the Kutta condition. In agreement with such concepts, vorticity could exist in a non-viscous medium, according to the prevailing consensus that the potential flow solutions characterize inviscid flows *per se*. This aforementioned assertion constitutes a historical debate concerning the fundamentals of fluid mechanics, related to the viscous [3], [4] or non-viscous [5], [6], [7] origin of vorticity, which is directly related to the generation of fluid force, including its perpendicular projection (lift). A discussion of this topic is beyond the scope of the present research, and ideally, it should be approached from theoretical and experimental [8] perspectives, rather than numerical ones. In this regard, it is easy to demonstrate that, when implementing the finite volume method (FVM) to solve the Euler equations, simulating an inviscid flow past a body yields a vorticity field (and force), regardless of the selected boundary condition on the surface (free slip or non-slip wall). Needless to say, this result seems to contradict the widely accepted wisdom that vorticity cannot be created in a non-viscous medium. However, numerical dissipation (by artificial viscosity) and discretization errors which could result in a lack of conservation of flow properties, are more common justifications for such behavior.

Whatever the true origin of vorticity may be, the results in this manuscript demonstrate that the FTVM redistributes detached circulation-vorticity more effectively than an attached flow scheme, which is typical of current potential flow solutions. This is true for both partially (pre-stall) and massively (post-stall) separated fluid flow conditions. In this sense, the reasoning and demonstrations presented in a recent, widely disseminated publication [9] are compatible with the idea of attributing viscosity to potential flow solutions intrinsically. In other words, they provide a viscous justification for the origin of circulation and lift within the scope of the PFT. This would explain the perfect agreement between the ideal (inviscid) and physical (viscous) results for the lift coefficient (CL) in lifting bodies at low angles of attack (AoA) without any viscous BL correction [10], [11]. In any case, if the origin of vorticity is assumed to be inviscid, the FTVM should be solving for a general purpose, including inviscid and viscous contributions. The level of fluid attachment or detachment will be determined, mainly in the low-Re range, by the viscous diffusion method (e.g., core spreading method; CSM). High-fidelity simulations (to capture the fine grain details within the BL, hence a precise solution) and a more in-depth analysis could confirm this later. On the other hand, if the origin of vorticity is assumed to be viscous, the FTVM is approximating, ideally, only for massively fluid detachment (or assumed *fully* inviscid or *very* high-Re) cases. However it would not be solving shear stresses within the viscous BL. In such a case, an additional scheme based on shear stress calculation [12] must be included to account precisely for the viscous effect near the solid surface; *away* from the wall, the vorticity diffusion due to fluid viscosity is solved by the implemented viscous scheme (e.g., CSM). Regardless of the aforementioned, the main objective of the current research is achieved: to develop a three-dimensional numerical method for solving fluid separation past thin bodies; its extension to thick ones is an ongoing research project.

Theoretically, the principles underlying the vortex methods (VMs) are inherently linked to the vortex panel and vortex lattice approaches. This theoretical framework makes it possible to explain fluid motion through a Lagrangian or meshless description. This eliminates the need for turbulence models to determine regions of fluid separation on the surface [13]. It has been demonstrated that VMs can solve both homogeneous isotropic and shear turbulence with a higher degree of accuracy when compared with pseudo-spectral DNS and finite difference methods [14]. This finding indicates that VMs have the potential to provide a viable solution for more complex fluid dynamics simulations. The efficacy of these results extends beyond the realm of engineering applications, rendering them a valuable resource for fundamental research. Furthermore, VMs are a valuable asset for the field of CFD due to their potential for enhanced efficiency through the use of modern and future hardware, as well as more advanced numerical implementations.

## 2 Methods

The incompressible Navier-Stokes equations (i-NSE) are a set of nonlinear partial differential equations (PDEs) that approximate the numerical solution for viscous fluid dynamics as it can include forces such as gravity, pressure, viscous diffusion, and advection. These equations are generally described in their velocity-pressure ($v - p$) formulation; however, they can be transformed into their velocity-vorticity ($v - \omega$) form [15], where the pressure term disappears after vectorial manipulations:

$$\frac{\partial \vec{\omega}}{\partial t} + (\vec{u} \cdot \nabla)\vec{\omega} = (\vec{\omega} \cdot \nabla)\vec{u} + \nu \nabla^2 \vec{\omega} \tag{1}$$

On the left-hand side, the first term corresponds to the partial temporal variation of the (vectorial) vorticity, and the second term represents the advection and tilting of such a variable. The whole left-hand side of eq. (1) can be interpreted as the total temporal variation of vectorial vorticity ($D\vec{\omega}/Dt$). The first term on the right-hand side corresponds to vortex squeezing-stretching, and the second term represents diffusion of vorticity due to fluid viscosity, or viscous diffusion. Removing the viscous term from such equations reduces the i-NSE to the i-Euler equations, which describe incompressible, inviscid, and rotational ($\omega \neq 0$) flow. This rotational condition is a well-known characteristic of these simplified equations of motion. Note that in the context of a non-viscous origin of vorticity theory, the presence of a vorticity field around an arbitrary body can be justified.

So, the total temporal variation of vorticity is obtained by eq. (2).

$$\left.\frac{D\vec{\omega}}{Dt}\right|_{c,t+\Delta t} = \frac{|\vec{\Gamma}|}{V_{c,t}} \left.\frac{D\overrightarrow{\delta L}}{Dt}\right|_{t+\Delta t} \quad (2)$$

where $|\vec{\Gamma}|$ is the magnitude of the vectorial circulation of the vortex tube, $V_{c,t}$ is its corresponding cylindrical (subscript $c$) volume at the previous time, and $D\overrightarrow{\delta L}/Dt$ is the total temporal variation of the tilting vector ($\overrightarrow{\delta L}$) after the advection step implementing a trajectory integration scheme (e.g., second-order Adams-Bashforth).

To avoid being repetitive, the numerical methodology followed, a finite difference-based scheme, has been extensively described and explained in the previous corresponding publication [16]. This methodology is in contrast to the vortex particle method (VPM) [17], which follows a different approach consisting of the temporal variation of circulation instead of vorticity. The FTVM includes precise calculations of vortex tilting, squeezing, and stretching by conserving total circulation and vorticity (by varying vortex volumes) at each time-step to maintain a stable solution throughout the simulation. It should be emphasized that all of the original FTVM concepts have been applied to the post-stall condition in the same way that they were applied to the original pre-stall calculation, except for the plate's leading edge wake; during the generation of a new layer of vortices at each iteration, an intuitive sign rotation is given (positive according to the bound circulation; see Fig. 1). Doing it this way ensures that all the external edges (trailing, lateral, and leading) are force-free within the scope of the KJ force calculation, since the circulation-vorticity is detached downstream with the correct orientation. As is standard in the context of FTVM, internal detached vorticity is obtained by the difference in circulation between each pair of bound vortex/vorton rings. The visual results show that well-defined *wingtip* vortices appeared in the pre-stall condition, while a characteristic turbulent behavior was observed in the post-stall case.

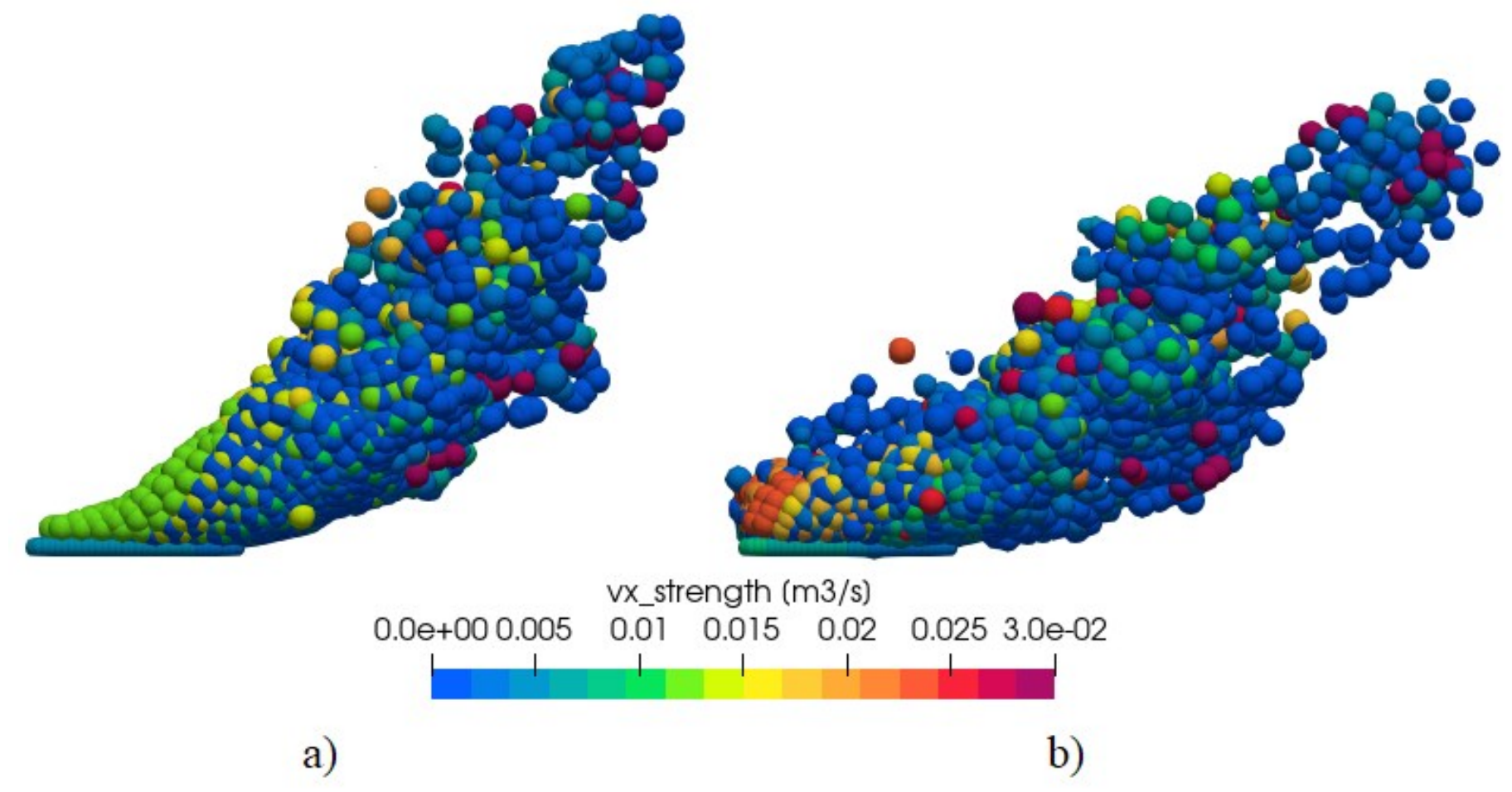

**Fig. 1** Vortex wake past a quadrangular flat plate (10×10) at $\alpha = 45°$ after 25 iterations in a) pre-stall and b) post-stall conditions (lateral view).

## 2.1 Viscous diffusion: the core spreading method (CSM)

The CSM [18] is one of the simplest schemes for accounting for vorticity diffusion due to fluid viscosity. It allows each vortex volume to increase over time, resulting in a variation in its induced velocity. Implementing the CSM in the FTVM involves increasing the vortex core radius (v.c.r.) of a vortex tube represented by a cylinder ($c$).

This is achieved by multiplying the kinematic viscosity value ($\nu$) by a constant ($k$) and then dividing it by the current v.c.r. ($\sigma$) of the vortex tube:

$$\left.\frac{d\sigma}{dt}\right|_{c,t+\Delta t}^{viscous} = \frac{k\nu}{\sigma_{c,t}} \tag{3}$$

In this work, the constant value is set to 1 ($k = 1$) since it corresponds to the implemented three-dimensional regularization function, a Gaussian error function (*erf*). Although the CSM originates from a mathematically exact solution of the diffusion equation, it cannot be extended to all simulation scenarios. For example, it cannot be used when fluid viscosity values are high because vortex volumes grow quickly, which causes non-physical conditions, such as vorticity penetration into or across the body. In such cases, an additional technique based on splitting vorticity must be implemented to divide a parent vortex element into a set of child vortices [19]. While this moderately increases code complexity, it ensures an appropriate numerical solution [20].

### 2.2 Force calculation: Andronov-Guvernyuk-Dynnikova's (AGD) method

The KJ is a well-proven implementation that allows to obtain the perpendicular projection (lift) of the aerodynamic force and the drag due to lift (induced drag) under the assumption of a fully-attached circulation scheme to the surface, which is typical of potential flow solutions. Such a methodology has been explained and implemented in the previous part of this research [16], where detaching circulation-vorticity from the entire surface of flat plates allowed an additional part of the pressure drag component (form drag) to be obtained, greatly improving the calculation.

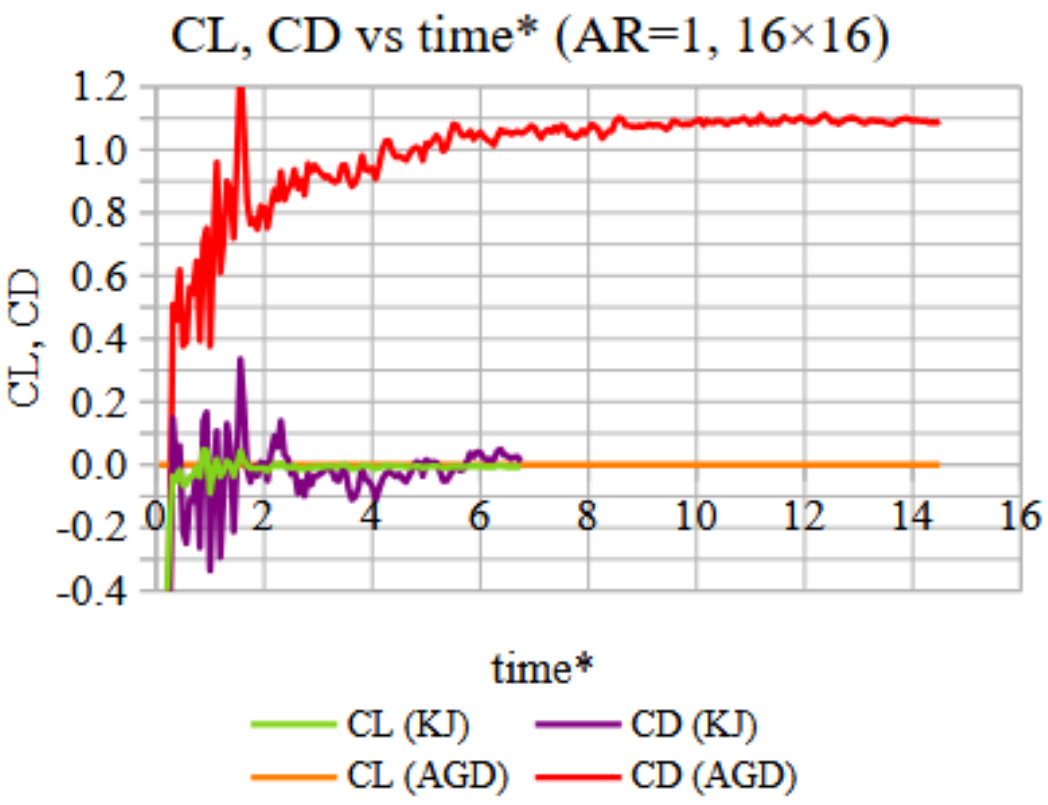

**Fig. 2** Lift and drag coefficients vs non-dimensional time for an $AR = 1$ flat plate at $\alpha = 90°$ (KJ vs AGD methods).

Figure 2 shows a comparison of an $AR = 1$ flat plate (16×16) at a perpendicular flow condition ($\alpha = 90°$) between the KJ and an alternative force calculation method [21] called here as *the AGD method* (due to the last names of the authors) that is well-tested and previously published. This method is based on normal pressure calculations. It is clear from these results that the KJ calculation cannot produce a practical result for the drag component since it was originally conceived from lifting surface concepts; its average value approaches zero over time. On the other hand, the AGD method yields a positive, well-defined value (in the range of a unit) for CD once the steady-state solution is reached. Validating it is the core of this research. The AGD method has been successfully applied to two-dimensional (viscous vortex domains; VVD [22], [23]) and three-dimensional (vortons [24] and vortex loops [25]) VMs.

The non-dimensional normal pressure ($p$) acting on each control point is obtained by considering a reference far-field pressure ($p_\infty$) and four velocity-related terms:

$$p(\vec{r},t) = p_\infty + \rho_\infty \left[ \frac{\left|\vec{V}_\infty\right|^2}{2} - \frac{\left|\vec{V}(\vec{r},t)\right|^2}{2} - \frac{1}{4\pi\Delta t} \sum_{i=1}^{N} A_i\left(\Gamma_i - \Gamma_i^{\,*}\right) + \sum_{k=1}^{K} \vec{v}_k(\vec{r}) \cdot \vec{V}(\vec{r}_k t) \right] \tag{4}$$

The first term corresponds to the free-stream velocity ($\vec{V}_\infty$). The second term corresponds to the induced velocity caused by all detached vorticity and is obtained by:

$$\vec{V}(\vec{r},t) = \vec{V}_\infty + \sum_{k=1}^{K} \vec{v}_k(\vec{r}) \tag{5}$$

where $\vec{v}_k(\vec{r})$ is the induced velocity by the k-vorton/blob on the control point. The third term involves the temporal variation of bound circulation ($\Gamma$) and the solid angle ($A$) projection of the i-discretized bounded vortex/vorton ring (*panel*) on the control point. For the flat plate case, this value comes from the corresponding element in a diagonal matrix (zeros except at the diagonal). The last term involves the dot product of the induced velocity and the k-vorton/blob kinematic velocity ($\vec{V}(\vec{r}_k t)$) vectors, the latter calculated as the mean vectorial velocity of the vortex tube endpoints' velocities.

Once the normal pressure has been obtained by eq. (4), the half of the total force acting on the infinitely thin plate is obtained by summing over *N panels*:

$$\vec{F}(t) = -\sum_{i=1}^{N} p(\vec{r}_i,t)\, s_i \cdot \hat{n}_i \tag{6}$$

where $s_i$ is the area of the *i-panel* and $\hat{n}_i$ is its corresponding i-unit normal vector. Since the current implementation of the FTVM solves for fluid flow past an infinitely thin (zero-thickness) flat plate, the calculated acting force is only applied to the upper side of the plate (at one control point with a positive unit normal vector; $+\hat{n}_i$), while the lower side (with a negative unit normal vector; $-\hat{n}_i$) is not considered.

Thus, to determine the total force acting on both sides of the plate, the pressure on the *panel* (or the total vector force) must be duplicated:

$$p_{u-d}(\vec{r}_i, t) = -2p(\vec{r}_i, t) \quad (7)$$

or,

$$\overrightarrow{F_T}(t) = -2\vec{F}(t) \quad (8)$$

To avoid this additional operation, a double-layer scheme should be implemented to remain consistent with the original (on a thick body). Such a complex methodology requires releasing detached vorticity from the top and bottom of each thin *panel*, which would increase computational costs. The single-layer scheme presented in this research should be understood as a simplified version of that general case.

As usual, the total force is obtained in the body's reference frame and must be transformed into the wind's reference frame to calculate the aerodynamic coefficients (lift and drag; CL, and CD, respectively):

$$CL = C_z\cos(\alpha) - C_x\sin(\alpha) \quad (9)$$

$$CD = C_z\sin(\alpha) + C_x\cos(\alpha) \quad (10)$$

where $\alpha$ is the angle of attack (AoA), $C_x$ and $C_z$ are the projections of the force coefficient in the tangential and normal directions, respectively. The normal coefficient (CN) of the total force is obtained by:

$$CN = CL\cos(\alpha) + CD\sin(\alpha) \quad (11)$$

## 3 Results

The results obtained mainly correspond to a quadrangular flat plate with an aspect ratio of 1 ($AR = 1$) at four angles of attack ($\alpha = 45°$, $\alpha = 60°$, $\alpha = 75°$, and $\alpha = 90°$ with 15° intervals) in the post-stall condition. As expected, longer simulations are required to reach a steady-state, if one exists, or a sufficiently developed unsteady wake. The current version of the *VortoNeX code* (v1.1) [26] does not yet implement fast computation algorithms, particularly those that accelerate the calculation of induced velocities. Examples include the fast multipole method (FMM) and the Barnes-Hut method, which reduce the computational complexity from $O(n^2)$ to $O(n)$ and $O(n\log n)$, respectively. All simulations were performed on an Intel Core i9 processor at 3.20 GHz with 32 GB of RAM and 22 out of 24 available threads (as opposed to 8 threads in [16]). The longest simulation took around 30 wall-clock hours in parallel mode.

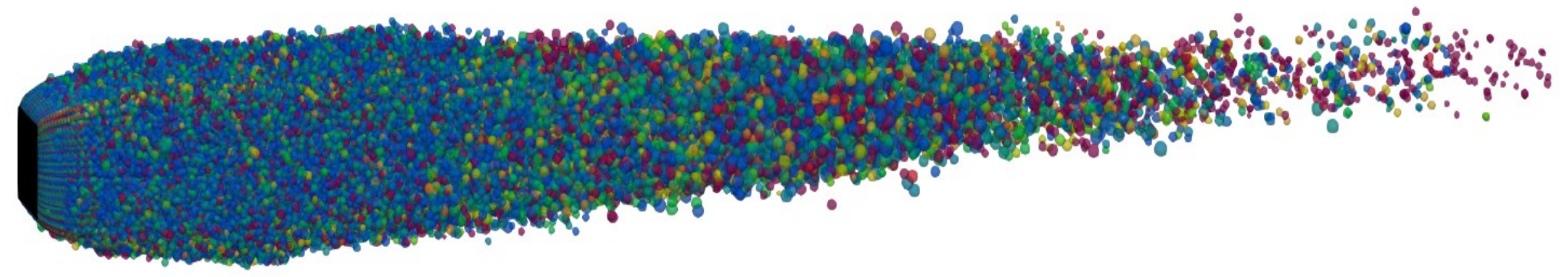

**Fig. 3** Vortex wake past a quadrangular ($AR = 1$) flat plate at $t^* = 14.5$ ($Re = 210{,}000$; approx. 60,000 detached vortons/blobs).

### 3.1 Input parameters

To maintain consistency, 3 of the 5 main input parameters were selected the same as they were in the corresponding validation for the pre-stall condition (16×16 discretized elements) in [16]. The kinematic fluid viscosity value ($\nu = 0.00000476$) was chosen based on $Re = 210{,}000$ from the reference experimental data used for validation [27]. The numerical values for the 5 main input parameters are shown in Table 1. The remaining parameters are set to unity: flow density ($\rho$), free-stream velocity magnitude ($q_\infty$), first wake row length factor ($\phi$), and reference pressure ($p_\infty$) for force calculation.

**Table 1** Main input parameters for all simulation cases for the $AR = 1$ configuration.

| Parameter | Numerical value |
|---|---|
| $\Delta t$ | 0.0625 |
| $\sigma_0$ | 0.0442 |
| $\varepsilon$ | 0.0442 |
| $\Phi$ | 0.01 |
| $\nu$ | $4.76\times10^{-6}$ |

As in previous related work, the time-step ($\Delta t$) is selected so that the chord length ($c = 1$) is divided by the number of discretized elements (16) in the chordwise direction. This results in a Courant–Friedrich–Lewy number of one ($CFL = \Delta x/\Delta t = 1$). The initial vortex core radius ($\sigma_0$) and the normal distance from the surface ($\varepsilon$) for the layer of nascent vortons are set to the same value ($\sigma_0 = \varepsilon = 0.0442$) in order to prevent a non-physical condition in which the vorticity crosses or penetrates the plate. The first one is obtained geometrically to ensure that the overlap between the discretized bound vortons is equal to one ($\lambda = 1$), as described in [16].

*3.1.1 Wake length (Φ) for the initial solution*

In the case of massively separated fluid flow (post-stall condition), the wake length ($\Phi$) used to calculate the initial solution (i.e., the circulation distribution on the plate at $t^* = 0$) was reduced by a factor of 100 ($\Phi = 0.01$), unlike in the case of partially separated fluid flow (pre-stall condition). This modification is based on a physical justification: for the pre-stall condition, the temporal change in fluid flow variables (e.g., circulation, vorticity, and velocity) is less abrupt than for the post-stall condition. This is because there are no longer smooth and well-defined vortex structures (e.g., wingtips) but turbulent behavior (see Fig. 1). For a steady-state simulation, for instance, a value in the order of 40 times the chord length ($\Phi = 40$), as in [28], can be selected since it is considered that the vortex wake has developed far downstream and reached a steady-state solution. On the other hand, using a unit value ($\Phi = 1$) as in [16] leads to instability in the time-marching solution during the first time-steps for the post-stall condition.

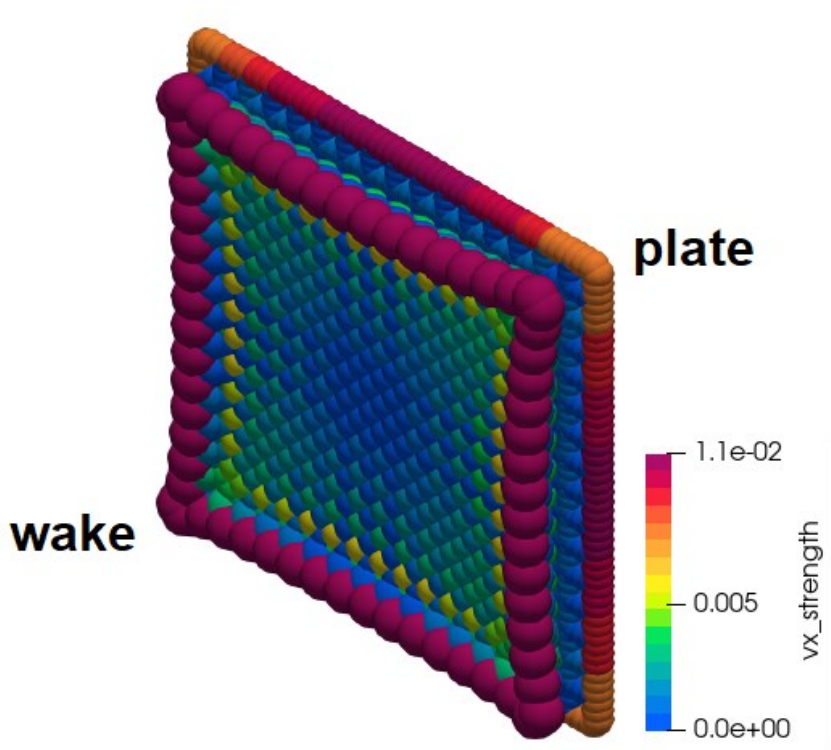

**Fig. 4** First vorton/blob layer on a normal quadrangular flat plate (16×16) after one iteration.

Figure 4 shows the vortex strength, or magnitude of circulation, for both the discretized flat plate and the vortex wake after the first iteration. Note that, despite the fact that this variable is concentrated along the edges of the plate, the interior values are not strictly zero.

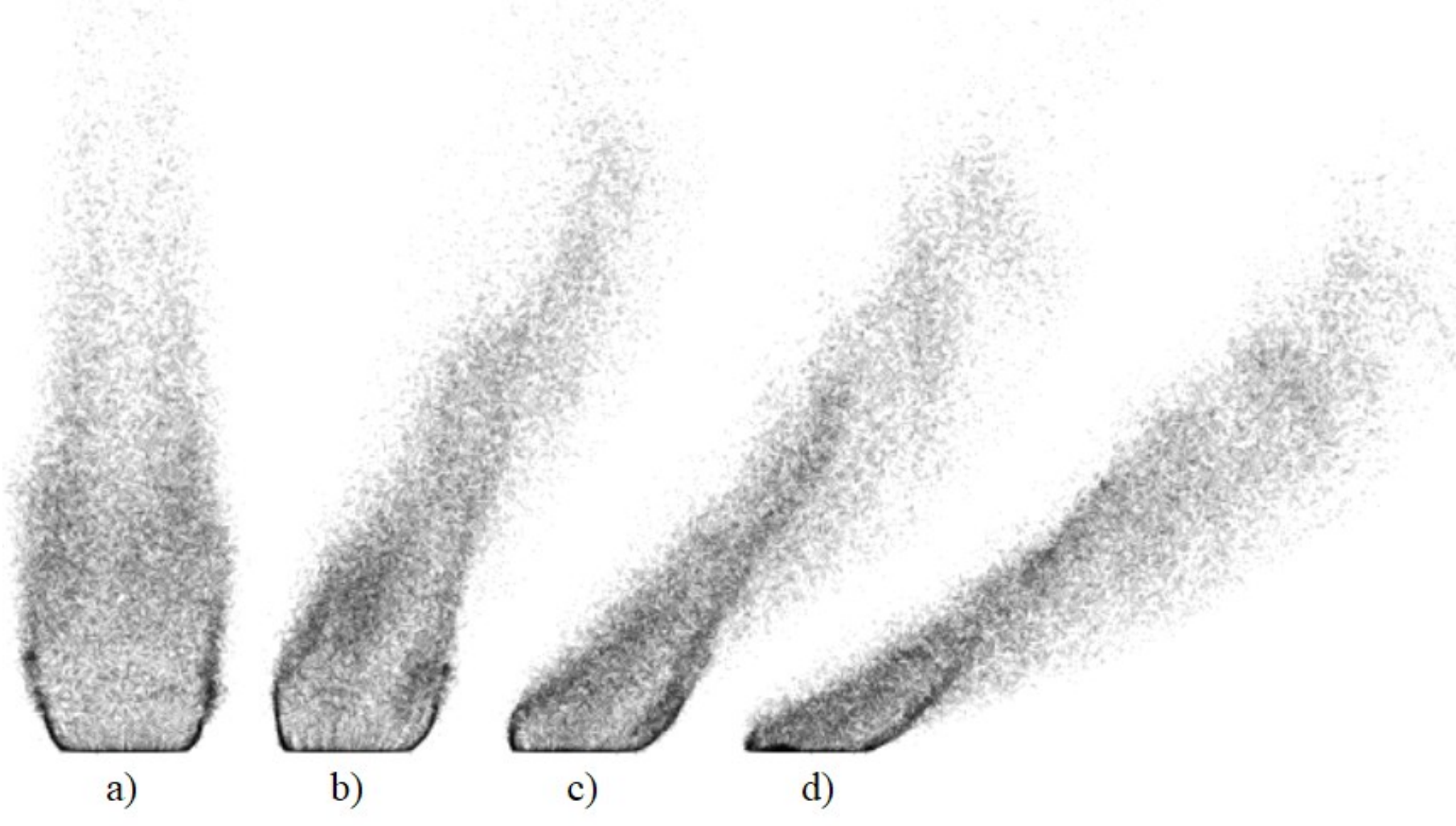

**Fig. 5** Wake patterns behind a quadrangular flat plate in a post-stall condition at different angles of attack (a) 90°, b) 75°, c) 60°, and d) 45°) at $t^* = 7$ (lateral view).

Figure 5 shows four vortex wake patterns at different AoAs, where each point represents a vortex center. The selected level of transparency (0.1) allows visualization of regions with a concentration of vorticity. The corresponding animation [29] shows vortex wake dynamics, including recirculation and turbulence.

### 3.2 Numerical results: verification and validation (v&v)

Since the current numerical implementation follows the same concepts verified and validated in previous related work [16] (except for the treatment of the leading edge), the verification phase relative to wake dynamics is assumed complete. As mentioned earlier, the only difference between the current and previous implementations concerns the calculation of the aerodynamic force.

#### *3.2.1 Unsteady analysis: CL, CD, and CN vs non-dimensional time*

In this section, six simulations were performed for the quadrangular flat plate case ($Re = 210{,}000$) at various inclinations. In addition to the previously mentioned AoAs, two more cases ($\alpha = 65°$ and $\alpha = 70°$) were included to extend the numerical analysis. The reference experimental data retrieved from [27] are shown by the dotted lines in the corresponding colors. As expected, some results correspond to stable (steady-state) solutions ($\alpha = 90°$ and $\alpha = 75°$), while others show oscillatory behavior ($\alpha = 45°$ and $\alpha = 60°$).

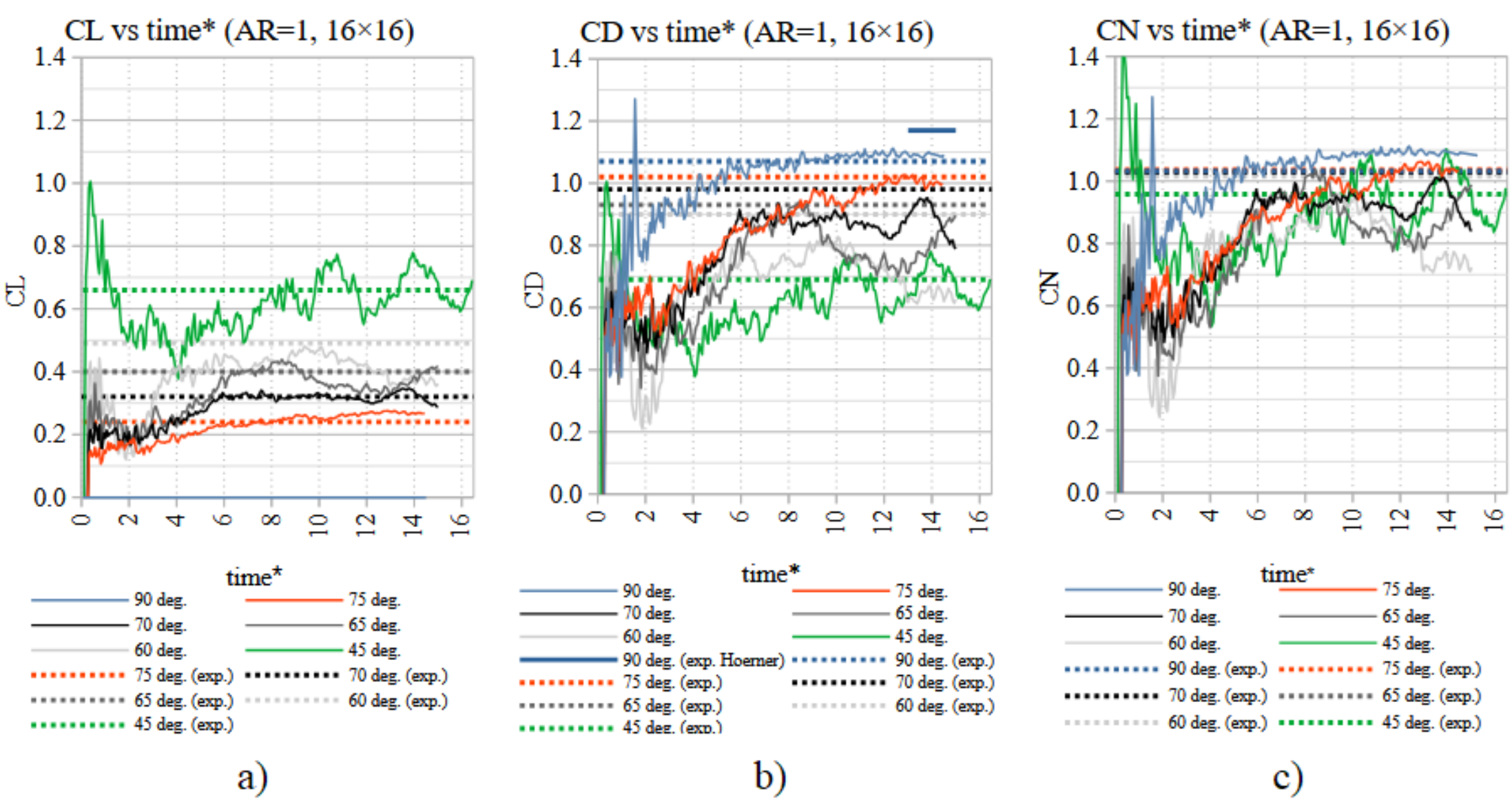

**Fig. 6** a) Lift, b) drag, and c) normal coefficients vs non-dimensional time for different AoAs.

Due to the nature and complexity of the numerical data, a qualitative analysis is performed here, with the quantitative analysis to follow. As observed, the lift coefficient (CL) is consistent across most of the simulations within the analyzed temporal range and oscillates around the mean experimental value. In this regard, the $\alpha = 45°$ case is of particular interest, as it exhibits a well-defined oscillatory solution after the initial transient phase. On the other hand, the drag coefficient (CD) can be determined with sufficient precision once a settling time is reached, particularly for $\alpha = 90°$, $\alpha = 75°$, and $\alpha = 45°$. The remaining cases seem to be underestimated compared to the reference experimental data. However, it should be noted that slight differences between the numerical and experimental drag results could be caused by the thickness factor. The numerical simulation is performed on an infinitely thin plate, while the physical plate has a thickness of around 4% (with straight edges) [27]. A more precise numerical analysis in the future could confirm this justification. The numerical CD calculations exhibit behavior similar to that of the normal coefficient (CN), indicating slightly underestimated mean values compared to the reference data. However, the $\alpha = 45°$ case matches the expected result. Note that a discrepancy of around 5% appears for the perpendicular flow case ($\alpha = 90°$). Ideally, the CN and CD values would be the same ($CN = CD = 1.07$). However, the former experimental value is underestimated ($CN = 1.03$), which likely indicates an issue with physical measurement. Consequently, the numerical CN is overestimated ($CN = 1.08$) by practically the same percentage.

Regarding the reference experimental value for the CD of a quadrangular flat plate normal to the flow, the classic value shown in most literature ($CD = 1.17$) is based on [30], which is 9% higher than the value in [27]. As with most physical tests, uncertainties can arise from various factors, such as the level of turbulence in the free-stream, load cell and mechanical balance calibration, among others. Therefore, any reference value must be carefully considered. For this reason, the reference used to validate the current research [27] was selected, as it shows experimental values across the entire post-stall range, not just for a single AoA. Consequently, all results, including numerical ones, must be interpreted with caution.

*3.2.2 Vortex stretching: variable vs constant vortex volumes*

Two additional simulations were performed for the cases of $\alpha = 45°$ and $\alpha = 90°$, which consist of maintaining a constant vortex volume for the detached vortons (after the advection and vortex stretching steps). This strictly violates the conservation of circulation between each time-step (see equations 2–5 in [16]). Note that even with a constant volume scheme, each vorton/blob grows due to viscous diffusion by the CSM (they only remain strictly constant in the inviscid case, when $\nu = 0$). The objective of these additional simulations is to qualitatively understand how precise vortex stretching calculations that strictly conserve total circulation and vorticity can improve a faster, simpler constant volume solution.

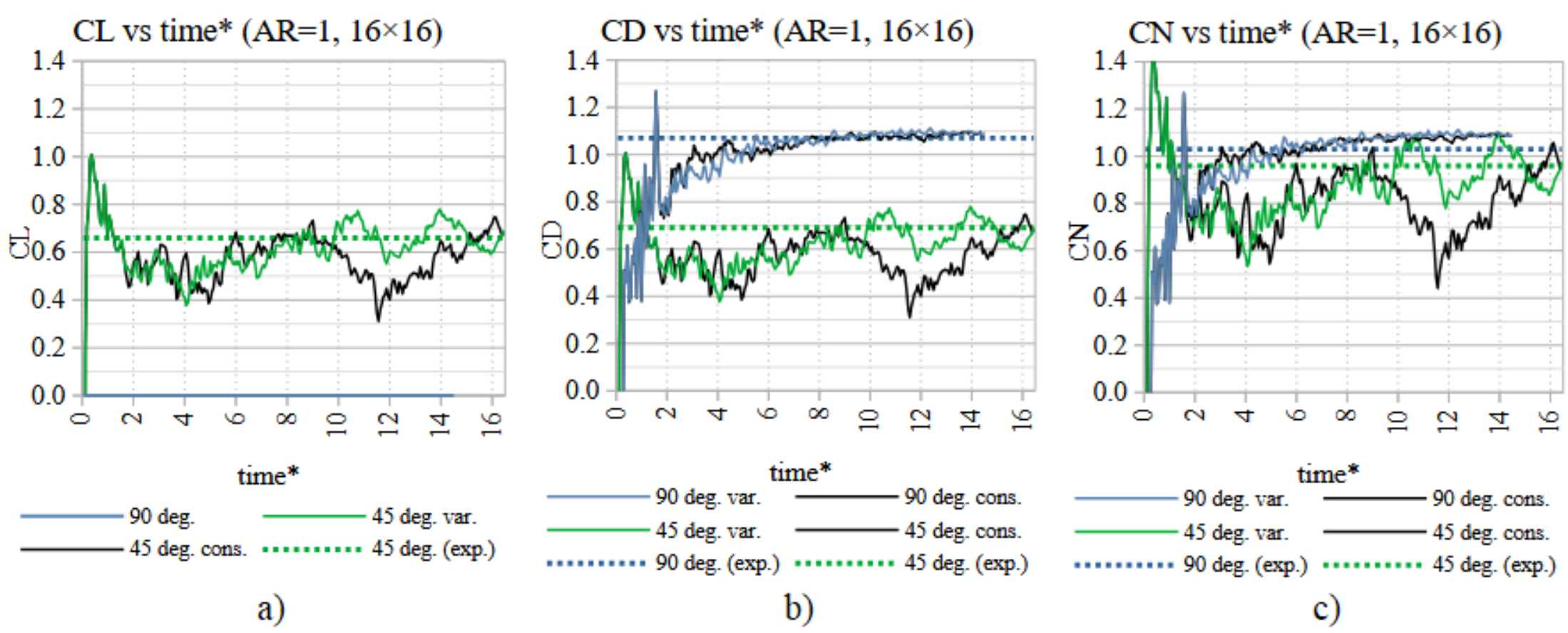

**Fig. 7** a) Lift, b) drag, and c) normal coefficients vs non-dimensional time for the cases of $\alpha = 45°$ and $\alpha = 90°$ (variable and constant vortex volumes).

For the perpendicular flow case ($\alpha = 90°$), it is clear that the constant volume scheme exhibits behavior quite similar to that of the variable volume scheme. In fact, the two schemes reach nearly identical steady-state values at the end of the simulations ($CD = CN = 1.08$ and $CD = CN = 1.07$ for the variable and constant volume solutions, respectively). However, for the inclined case ($\alpha = 45°$), the variable vortex volume scheme appears to offer more precise results by oscillating around the expected reference values ($CD = 0.69$ and $CN = 0.96$). In contrast, the constant volume scheme exhibits higher amplitude and more erratic behavior. A longer simulation or additional reference experimental data could help determine which better approximates turbulent behavior. A more precise analysis must be conducted in the future (i.e., turbulence analysis).

**Table 2** Numerical and reference experimental data for all simulations performed.

| | | Numerical (avg.) | | | Experimental | | | Difference (%) | | | Range avg. |
|---|---|---|---|---|---|---|---|---|---|---|---|
| | | *CL* | *CD* | *CN* | *CL* | *CD* | *CN* | *CL* | *CD* | *CN* | $t^*$ |
| AoA [°] | 45 | 0.6641 | 0.6641 | 0.9392 | 0.66 | 0.69 | 0.96 | 0.41 | -3.98 | -2.01 | 9.0-16.0 |
| | 45-c | 0.5577 | 0.5577 | 0.7887 | 0.66 | 0.69 | 0.96 | -15.50 | -19.17 | -17.84 | 9.0-16.0 |
| | 60 | 0.4042 | 0.7001 | 0.8084 | 0.49 | 0.90 | 1.02 | -17.40 | -21.97 | -20.40 | 10.0-15.0 |
| | 65 | 0.3781 | 0.8108 | 0.8947 | 0.40 | 0.93 | 1.03 | -5.48 | -12.81 | -13.14 | 7.0-15.0 |
| | 70 | 0.3195 | 0.8779 | 0.9342 | 0.32 | 0.98 | 1.03 | -0.15 | -10.42 | -9.30 | 6.0-15.0 |
| | 75 | 0.2681 | 1.0005 | 1.0358 | 0.24 | 1.02 | 1.04 | 10.89 | -1.48 | -0.05 | 11.0-14.0 |
| | 90 | 0.0000 | 1.0824 | 1.0824 | 0.00 | 1.07 | 1.03 | 0.00 | 1.00 | 4.81 | 7.0-14.0 |
| | 90-c | 0.0000 | 1.0747 | 1.0747 | 0.00 | 1.07 | 1.03 | 1.00 | 0.44 | 4.34 | 7.0-14.0 |

Table 2 shows a comparison of all the numerical and experimental data. Repeated AoA cases correspond to constant (-c) vortex volume calculations. The range for averaging the numerical values was chosen according to the following criteria: the lower limit was selected when the numerical solution reached the expected reference value after an initial transient phase, and the upper limit was selected as the maximum possible value (both rounded to the nearest integer of non-dimensional time). For a better visualization of these data, they are presented graphically below.

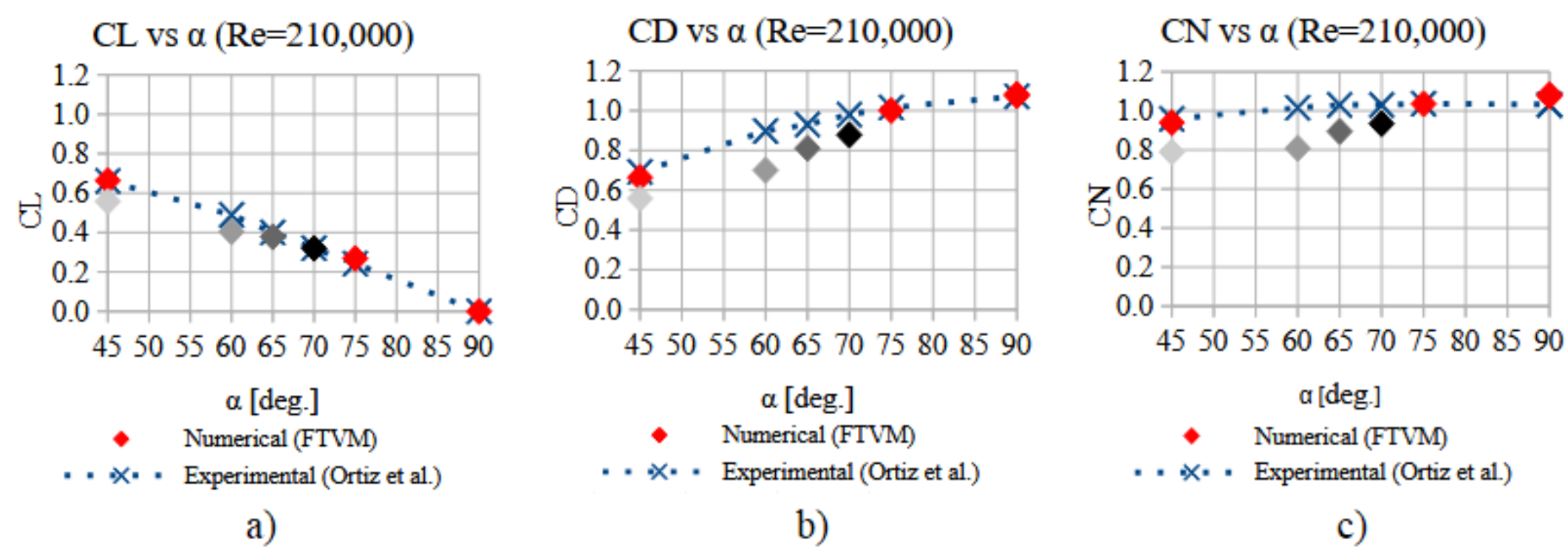

**Fig. 8** Time-averaged a) lift, b) drag, and c) normal coefficients vs AoA for a quadrangular flat plate at $Re = 210{,}000$.

Figure 8 shows the time-averaged lift, drag, and normal coefficients corresponding to the numerical data in Table 2. The red markers closely match the reference experimental data (with an error of less than 5%, except for CL in the $\alpha = 75°$ case), while the gray markers underestimate the data by around 20% (for CD and CN). As previously mentioned, these cases correspond to oscillatory solutions, which are highly turbulent, and require further analysis. Such research should consider the aforementioned thickness factor. Additionally, longer simulations may be necessary for correctly interpreting such numerical results. Interestingly, a well-defined trend is observed for the unconverged cases ($\alpha = 60°$, $\alpha = 65°$, and $\alpha = 70°$) since they tend to reach the expected reference value as the AoA increases. The missing analyzed cases ($\alpha = 50°$ and $\alpha = 55°$) could show an inverse trend. The lighter gray markers show additional results for the constant vortex volume scheme ($\alpha = 45°$ and $\alpha = 90°$ cases). For the perpendicular flow case, the lighter gray marker appears hidden behind the red marker.

### 3.3 Additional simulation: perpendicular $AR = 0.5$ flat plate case

An additional simulation was performed for a rectangular flat plate ($8\times16$) oriented normal to the free-stream ($\alpha = 90°$). All input parameters were the same as for the $AR = 1$ plate case except for the fluid viscosity, which was set to $\nu = 0.00000666$ for $Re = 150{,}000$. There are few formal publications or reported data related to low AR aerodynamics for flat plates in the pre-stall condition, and even fewer for the post-stall condition. The number is significantly reduced for a rectangular $AR = 0.5$ (and quadrangular $AR = 1$) flat plate. However, as expected, an $AR = 2$ flat plate normal to the flow will have the same CD value for an $AR = 0.5$ rectangular configuration. Relatively recent publications on perpendicular $AR = 0.5$ and $AR = 2$ flat plates are shown in Table 3.

**Table 3** Recent publications for flat plates that include results for the $\alpha = 90°$ case.

| AR | CD | Re | Method | Reference | Year |
|---|---|---|---|---|---|
| 2 | 1.51 | 100 | Experimental (oil tank) | Taira, K. et al. [31] | 2009 |
| 2 | 1.49 | | Numerical (Immersed BM) | | |
| 2 | 1.50 | 20,000 | Experimental (water tunnel) | Granlund, K. et al. [32] | 2011 |
| 0.4 | 1.07 | 118,000 | Experimental (wind tunnel) | Ortiz, X. et al. [27] | 2015 |
| 0.6 | | 178,000 | | | |
| 0.5 | 0.90 | 150,000 | Numerical (LES) | Shademan, M. et al. [33] | 2020 |
| 2 | 0.94 | | | | |

Except for the first publication [31], all the others show experimental [32], [27] or numerical [33] results in the moderate-to-high Re range ($Re = 20{,}000$ to $Re = 178{,}000$). In the context of the present research, only the latter should be considered justified reference values, despite the fact that it is well known that Re has minimal physical implications for the steady-state CD value in conditions of massively separated (or inviscid-dominated) fluid flow, especially in flat plate configurations with sharp or flat edges.

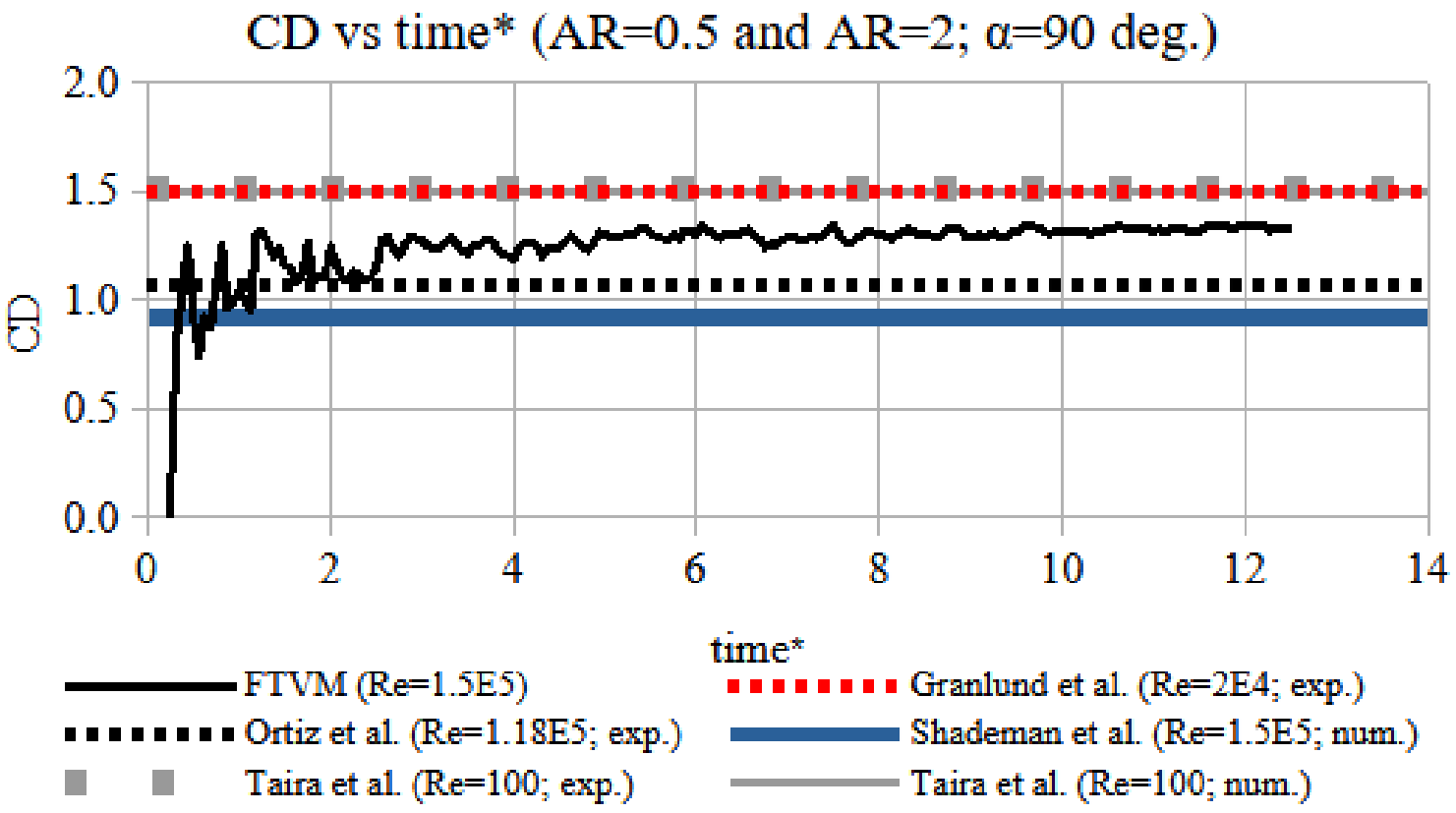

**Fig. 9** Drag coefficient vs non-dimensional time, along with reference steady-state values, for two flat plate configurations at normal flow ($\alpha = 90°$).

Figure 9 shows that, disregarding the low-Re case and regardless of the steady-state numerical value achieved by the FTVM ($CD = 1.31$), there is no correlation between previously reported experimental and numerical CD values, even for physical tests results, where the difference between such values is approximately 30%. A deeper, more extensive numerical analysis could be carried out in the context of the FTVM when reliable reference data is available. This inconsistency was previously reported by the author in [16] for an $AR = 0.5$ flat plate in the pre-stall condition.

## 4 Conclusions

The numerical results achieved are evidently conclusive: The Full Nonlinear Vortex Tube-Vorton Method (FTVM) is capable of reliably approximating fluid flow past a thin body with massive fluid separation (i.e., a stalled flat plate) involving complex, unsteady wake patterns characteristic of turbulent behavior. Such numerical results depend on a reduced set of input parameters, most of them related to temporal ($\Delta t$) and spatial ($\sigma_0$, $\varepsilon$, and $\Phi$) discretizations. This approach avoids the need for semi-empirical modeling (e.g., turbulence models) or convenient corrections for approximating fluid separation regions. As previously stated in the initial publication related to this research (*The Full Multiwake Vortex Lattice Method*; FMVLM) [28], vortex wake dynamics are governed by pure Lagrangian mechanics (and conservation laws for the incompressible case), regardless of its complexity. A correct distribution of circulation (hence, vorticity) on the body surface is the keypoint to achieving the current satisfactory results. This fact avoids dealing with concepts such as surrounding pressure distributions as a cause of fluid motion or absurd assumptions (e.g., perfectly attached flows in the context of the PFT) for simulating partially or massively fluid flow separation cases.

At this stage of development, the FTVM has been naturally extended to solve for fluid flow past a thick body (a sphere at $Re = 15{,}000$), yielding preliminary, satisfactory results (a precise steady-state CD). This follows the achievements in [25], including the well-resolved drag crisis phenomenon (by another vortex method). This extension helps to explain low-speed aerodynamics in a simple and straightforward manner. It also has the potential to address more complex fluid dynamics simulations from a vorticity-based perspective as this numerical method continues to evolve.

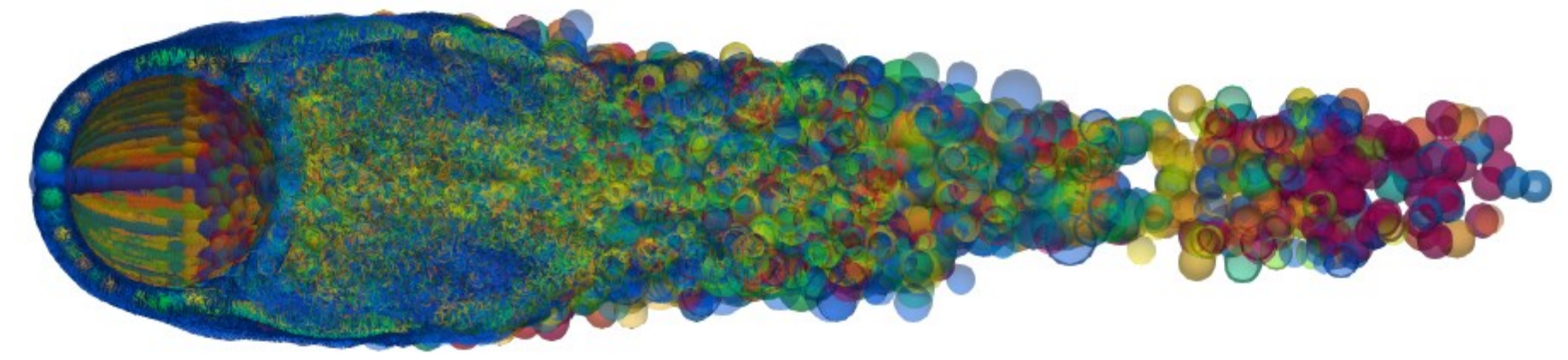

**Fig. 10** Vortex wake behind a sphere by the FTVM (200 surface elements; 300 detached vortons/blobs at each iteration; ongoing research).

**List of abbreviations**

AGD — Andronov-Guvernyuk-Dynnikova (calculation)
AoA — angle of attack
AR — aspect ratio
BL — boundary layer
CFD — computational fluid dynamics
CFL — Courant-Friedrich-Lewy (number)
CL, CD, CN — lift, drag, and normal coefficient
CSM — core spreading method
DNS — direct numerical simulation
FMM — fast multipole method
FMVLM — full multiwake vortex lattice method
FTVM — full nonlinear vortex tube-vorton method
FVM — finite volume method

GPU graphics processing unit

KJ Kutta-Zhukovski (calculation)

LES large eddy simulation

NSE Navier-Stokes equations

PDE partial differential equations

PFT potential flow theory

RANS Reynolds-averaged Navier-Stokes

Re Reynolds number

SST shear stress transport (model)

UFVLM unsteady full multiwake vortex lattice method

v.c.r. vortex core radius

VMs vortex methods

VPM vortex particle method

VVD viscous vortex domains (method)

## Declarations

**Acknowledgments**

Not applicable.

**Authors' contributions**

JCPG contributes all the computational related contents in this manuscript. LAJP analyzed and interpreted numerical results from an experimental and practical standpoint. All authors read and approved the final manuscript.

**Funding**

Not applicable.

**Availability of data and materials**

As with both previous codes related to this research, MultiVLM (FMVLM) and UMultiVLM (UFVLM), the corresponding open-source code, VortoNeX v1.1 (FTVM) [26], is available for distribution on www.github.com

**Competing interests**

The authors declare that they have no competing interests.

## Authors' information

JCPG is a PhD student on temporary leave of absence since October 2021. He has experience in Lagrangian mathematical modeling (parachute flight dynamics), linear stability analysis, computational fluid dynamics (by finite volume and lattice-Boltzmann methods) and programming vorticity-based methods, as well as experimental aerodynamics, including wind tunnel and drop tests of small-scale parachutes for different applications [34]. His early, formal research on vortex methods is protected by a PCT patent application [35].

LAJP earned his Master of Science degree from the National Polytechnic Institute (IPN), Mexico, in 2016. His master's research focused on the application of genetic algorithms to compressible aerodynamics of wings. Since then, he has been a faculty member at Tecnológico de Estudios Superiores de Ecatepec (TESE), where he teaches aerodynamics and related subjects. Additionally, he has served as the advisor for a Society of Automotive Engineers (SAE) AeroDesign team. Currently, his research involves the design and construction of unmanned aerial vehicles and flying wings.

## References

1. Spalart PR, Allmaras S (1992) A one-equation turbulence model for aerodynamic flows. Technical Report AIAA-92-0439. https://doi.org/10.2514/6.1992-439

2. Mohseni K, Xia, X (2016) Unsteady aerodynamics and trailing-edge vortex sheet of an airfoil. 54th AIAA Aerospace Sciences Meeting. https://arc.aiaa.org/doi/abs/10.2514/6.2016-1078

3. Wu J (1995) A theory of three-dimensional interfacial vorticity dynamics. Physics of Fluids 7, 2375–2395. https://doi.org/10.1063/1.868750

4. Wu J, Wu J (1993) Interactions between a solid surface and a viscous compressible flow field. Journal of Fluid Mechanics 254:183–211. https://doi.org/10.1017/S0022112093002083

5. Morton B (1984) The generation and decay of vorticity. Geophysical & Astrophysical Fluid Dynamics 28(3-4):277-308. https://doi.org/10.1080/03091928408230368

6. Morino L (1986) Helmholtz decomposition revisited: vorticity generation and trailing edge condition. Computational Mechanics 1:65–90. https://doi.org/10.1007/BF00298638

7. Terrington S, Hourigan K, Thompson M (2022) Vorticity generation and conservation on generalised interfaces in three-dimensional flows. Journal of Fluid Mechanics 936. https://doi.org/10.1017/jfm.2022.91

8. Craig PP, Pellam JR (1957) Observation of perfect potential flow in superfluid. Phys. Rev. 108, 1109. American Physical Society. https://link.aps.org/doi/10.1103/PhysRev.108.1109

9. Liu T (2021) Evolutionary understanding of airfoil lift. Advances in Aerodynamics 3(37). https://doi.org/10.1186/s42774-021-00089-4

10. Ozdemir YH, Barlas B (2021) 2D and 3D potential flow simulations around NACA 0012 with ground effect, Research Square. https://doi.org/10.21203/rs.3.rs-151154/v1

11. Karali H, Inalhan G, et al (2021) A new nonlinear lifting line method for aerodynamic analysis and deep learning modeling of small unmanned aerial vehicles. International Journal of Micro Air Vehicles 13. https://doi.org/10.1177/17568293211016817
12. Ojima A, Kamemoto K (2000) Numerical simulation of unsteady flow around three dimensional bluff bodies by an advanced vortex method. JSME International Journal Series B-fluids and Thermal Engineering 43:127-135. https://doi.org/10.1299/jsmeb.43.127
13. Spalart PR (1988) Vortex methods for separated flows. Tech. Rep., NASA, Moffett Field.
14. Yokota R, Obi S (2011) Vortex methods for the simulation of turbulent flows: Review, Journal of Fluid Science and Technology 6(1). https://doi.org/10.1299/jfst.6.14
15. Speziale VG (1987) On the advantages of the vorticity-velocity formulation of the equations of fluid dynamics. Journal of Computational Physics 73(2). https://doi.org/10.1016/0021-9991(87)90149-5
16. Pimentel JC (2024) The full non-linear vortex tube-vorton method: the pre-stall condition. Advances in Aerodynamics 6(13). https://doi.org/10.1186/s42774-023-00168-8
17. Winckelmans GS (1989) Topics in vortex methods for the computation of three- and two-dimensional incompressible unsteady flows. Dissertation, California Institute of Technology. https://doi.org/10.7907/19HD-DF80
18. Leonard A (1980) Vortex methods for flow simulation. J Comput Phys 37(3). https://doi.org/10.1016/0021-9991(80)90040-6
19. Zuhal LR, Dung DV, et al (2014) Core spreading vortex method for simulating 3D flow around bluff bodies. J Eng Technol Sci 46(4):436–454. https://doi.org/10.5614/j.eng.technol.sci.2014.46.4.7
20. Rossi LF (1996) Resurrecting core spreading vortex methods: a new scheme that is both deterministic and convergent. SIAM Journal on Scientific Computing 17(2):370–397. https://doi.org/10.1137/S10648275932543
21. Andronov PR, Guvernyuk SV, Dynnikova GY (2006) Vortex methods for calculation of non-stationary hydrodynamic loads (in Russian). Moscow State University. ISBN: 5-211-05256-0. https://istina.msu.ru/publications/book/1064403/
22. Dynnikov YA (2016) The Viscous Vortex Domains Method (in Russian). Github.com. https://github.com/vvflow/vvflow/blob/master/doc/documentation.pdf. Accessed 30 May 2025.
23. Marchevsky I, Sokol K, et al (2023) The VM2D open source code for two-dimensional incompressible flow simulation by using fully lagrangian vortex particle methods. Axioms 12(3):248. https://doi.org/10.3390/axioms12030248
24. Shcheglov GA (2008) On the application of vorton frames in the vortex particle method (in Russian). Bulletin of the Moscow State Technical University. https://vestniken.bmstu.ru/articles/255/eng/255.pdf. Accessed 30 May 2025.
25. Dergachev SA, Shcheglov GA (2018) Mathematical modeling of the hydrodynamic loading of an aircraft capsule using the vortex loop method (in Russian), Ph.D. dissertation. Moscow State Technical University. https://rusneb.ru/catalog/000199_000009_009858786/. Accessed

30 May 2025.

26. Pimentel JC (2025) VortoNeX v1.1 Fortran code. Github.com. https://github.com/CpimentelMx/VortoNeX. Accessed 30 May 2025.

27. Ortiz X, Rival D, Wood D (2015) Forces and moments on flat plates of small aspect ratio with application to PV wind loads and small wind turbine blades. Energies 8:2438-2453. https://doi.org/10.3390/en8042438

28. Pimentel JC (2023) The full multi-wake vortex lattice method: a detached flow model based on potential flow theory. Advances in Aerodynamics 5(22). http://doi.org/10.1186/s42774-023-00153-1

29. Pimentel JC (2025) 3D viscous fluid past a square flat plate in post-stall condition. Youtube.com. https://www.youtube.com/watch?v=JDerf6RRrLc. Accessed 30 May 2025.

30. Hoerner SF (1965) Fluid-Dynamic Drag, Bakersfield.

31. Taira K, Colonius T (2009) Three-dimensional flows around low-aspect-ratio flat-plate wings at low Reynolds numbers. Journal of Fluid Mechanics 623:187-207. https://doi.org/10.1017/S0022112008005314

32. Granlund KO, Ol MV, Bernal LP (2013) Unsteady pitching flat plates. Journal of Fluid Mechanics 733:R5. https://doi.org/10.1017/jfm.2013.444

33. Shademan M, Naghib-Lahouti A (2020) Effects of aspect ratio and inclination angle on aerodynamic loads of a flat plate. Advances in Aerodynamics 2(14). https://doi.org/10.1186/s42774-020-00038-7

34. Pimentel JC (2013) Personal brand website. https://www.chuteshiut.com. Accessed 30 May 2025.

35. Pimentel JC (2024) Full-surface detached vorticity method and system to solve fluid dynamics. World Intellectual Property Organization. PCT application: WO/2024/136634. https://patentscope.wipo.int/search/en/detail.jsf?docId=WO2024136634&_cid=P21-LXXE19-45988-1. Accessed 30 May 2025.